\documentclass[12pt]{iopart}
\usepackage{graphicx}
\usepackage{iopams}
\usepackage[colorlinks=true]{hyperref}

\def\tr{\,{\rm Tr\,}}\def\ttr{\,{\rm tr\,}}
\def\Ud{U^{\dagger}}
\def\text#1{{\rm #1}}
\def\bu{{\bi{u}}}\def\bv{{\bi{v}}}

\def\rO{{\rm O}}\def\rU{{\rm U}}
\def\({\left(}\def\){\right)}
\def\oh{\frac{1}{2}}
 \def\Ga{\alpha}\def\Gl{\lambda}\def\GO{\Omega}
\def\CM{{\cal M}}\def\CW{{\cal W}}\def\CZ{{\cal Z}}
\def\be{\begin{equation}}\def\ee{\end{equation}}
\def\diag{{\rm \,diag\,}}
\def\FU{F^{\rm(U)}}\def\FO{F^{\rm(O)}}
\def\ZU{Z^{\rm(U)}}\def\ZO{Z^{\rm(O)}}
\def\CZO{\CZ_{\rm O}}
\def\1oN{{1\over N}}
\def\ommit#1{{}}
\def\Ad{{A^\dagger}}\def\Bd{B^\dagger}

\def\pre#1{ ({\tt
#1})}

\begin{document}

\title[Large $N$ limit of integrals over the orthogonal group]{On the large N limit of matrix integrals over the orthogonal group}
\author{Jean-Bernard Zuber}
\address{LPTHE (CNRS, UMR 7589), Universit\'e Pierre et Marie Curie-Paris6, 75252 Paris Cedex, France.}
\ead{jean-bernard.zuber\,@\,upmc.fr}

\date{March 2008}

\begin{abstract}
\ommit{We reexamine the large $N$ limit of matrix integrals over the orthogonal group
O($N$) and their relation with those pertaining to the unitary group U($N$). 
We prove that $\lim_{N\to \infty}N^{-2}\int DO \exp N \tr JO $ is half the 
corresponding function in U($N$), and  a similar relation 
for  $\lim_{N\to \infty}\int DO \exp N \tr( A O B O^t ) $, for $A$ and $B$ 
both symmetric or both skew symmetric. }

The large $N$ limit of some matrix integrals over the orthogonal group
O($N$) and its relation with those pertaining to the unitary group U($N$)
are reexamined. 
It is proved that $\lim_{N\to \infty}N^{-2}\int DO \exp N \tr JO $ is half the 
corresponding function in U($N$), with  a similar relation 
for  $\lim_{N\to \infty}\int DO \exp N \tr( A O B O^t ) $, for $A$ and $B$ 
both symmetric or both skew symmetric. 

\end{abstract}

\maketitle

\section{Introduction}
Matrix integrals of the type
\be
\label{Theint} 
Z^{(G)}=\int_G D\Omega\, \rme^{ \kappa N \tr(A\Omega B\Omega^{-1})}\ ,
\ee
over a classical compact group $G=$ U($N$), O($N$) or Sp($N$),
with $\kappa$ a real parameter,  
appear frequently in theoretical
physics, from disordered systems \cite{Ka, MPR, review} to 2D quantum 
gravity \cite{DFGZJ} and related topics. They have also a mathematical 
interest in connection with integrability, statistics  and free probabilities.
They are sometimes called
matrix Bessel functions \cite{GuK}, or (generalized) HCIZ integrals.   
While the expression of $Z$ is well known for the group U($N$)
\cite{HC,IZ}, 
the situation with O($N$) is more subtle. The result is known for
{\it skew-symmetric} matrices $A$ and $B$  \cite{HC}, but its form is
only partially understood for the more frequently encountered case of
{\it symmetric} matrices, in spite of recent progress \cite{GuK,BH,BE}.

On the other hand, in the large $N$ limit, we expect things to
simplify \cite{We,BG}.
It is the purpose of this work to revisit this old problem 
and to show that 
$\log Z$ has  universality properties in the large $N$ limit, a 
pattern which does not seem to have been stressed enough before, 
at least in the physics literature,  see the historical note below.

The paper is organized as follows. In section 2, we discuss the related
but simpler case of the integral ``in an external field'',
\be 
\label{extfield}
\CZ_G=\int_G D\Omega\, \rme^{N \tr (J \Omega+ J^\dagger \Omega^\dagger)}\ 
\ee
(where the second term in the exponential will be omitted in the 
orthogonal case). We prove it enjoys a 
universality property in the large $N$ limit. In section 3, we turn to
the integral (\ref{Theint}), discuss its large $N$ limit and prove it
has a similar universality property. We also extend our integrals 
to cases where matrices $A$ and $B$ are neither symmetric (or Hermitian) 
nor antisymmetric.

\ommit{We shall make use of Dyson's label $\beta=1, 2, 4$ for  $G=$ U($N$), O($N$) or Sp($N$), respectiveley.}
For the sake of simplicity the present paper will be focused on the 
case of $G=$ O($N$) as compared to U($N$), but similar considerations
apply to Sp($N$). 


\section{The integral (\ref{extfield})}
\subsection{The basic integrals and their generating function }
Let us consider the basic integral over the orthogonal group O($N$)
\be 
\label{basic0}
{\cal I}_{\bi{i},\bi{j}}:=\int DO\, O_{i_1j_1}\cdots O_{i_{n}j_{n}}
\ee
where $O$ is an $N\times N$ orthogonal matrix and $\bi{i}$, resp.$\bi{j}$, are 
$n$-tuples of indices $i$ resp. $j$.

One may easily prove that ${\cal I}_{\bi{i},\bi{j}}$ is non vanishing only for even $n$ and has then the following general structure
\cite{CS1},
\be 
\label{basic}
\int DO\, O_{i_1j_1}\cdots O_{i_{2n}j_{2n}} =
\sum_{p_1, p_2\in P_{2n}} \delta^{p_1}_{\bf i} \delta^{p_2}_{\bf j} C(p_1,p_2)
\ee
where $P_{2n}$ denotes the set of pairings in $\{1,2,\cdots,2n\}$, 
\be
\!\!\!\!\!\!\!\!\!\!\!\!\!\!\!\!\!\!\!\!\!\!\!\!\!\!\!\!
\delta^{p}_{\bf i}=1 \quad {\rm if}\ \forall a, b,\ \ a=p(b) \Rightarrow i_a=i_b,
\quad i\in \{1,2,\cdots, N\}\  {\rm and}\ \delta^{p}_{\bf i}=0\ 
{\rm otherwise},
\ee
and where the coefficients $C(p_1,p_2)$ enjoy many properties and may 
be determined recursively, see Appendix A for a review.

The analogous basic integral for $\rU(N)$ is
\be 
\label{basicU}
\int DU\, U_{i_1j_1}\cdots U_{i_nj_n} \Ud_{k_1\ell_1} \cdots\Ud_{k_n\ell_n}=
\sum_{\tau,\sigma\in S_n} C([\sigma])
\prod_{a=1}^n \delta_{i_a\ell_{\tau(a)}} \delta_{j_a k_{\tau\sigma(a)}}\ ,
\ee
with a double sum over permutations $\sigma, \tau \in S_n$, the symmetric 
group, see Appendix A.

Another way of encoding these formulae is to use the generating functions
\be
{\cal Z_{\rm O}}(J,J^t;N)=\int DO\, \rme^{ N \tr J O} = \rme^{ \CW_{\rm O}(J.J^t,N)} \ee
\be{\cal Z_\rU}(J,J^\dagger;N)=\int DU \rme^{ N \tr (J U+ J^\dagger \Ud)}=
\rme^{ \CW_{\rm U}(J.J^\dagger,N)} \ ,
\ee
where $J$ is a generic ({\it i.e.} non symmetric, resp. non Hermitian) 
 matrix.  
$\CZ_\rO$ depends only on traces of powers of $J.J^t$, again by invariance of
the integral under $J\to O_1.J.O_2$. Likewise, $\CZ_\rU$ depends only 
on $J.J^\dagger$.  

\subsection{The large N limit  and its relation to the unitary case}\label{ASM}

In the large $N$ limit, one may show 
that 
\be
\label{WOinf}
W_{\rm O}(J.J^t) =\lim_{N\to\infty} N^{-2}\,\CW_{\rm O}(J.J^t,N)
\ee
 exists and is 
related to the corresponding expression for the unitary group.

Recall  the situation in the unitary case. The generating 
function  $\CZ_\rU$
has been studied extensively in the past. 
The limit $W_{\rm U}(J.J^\dagger)
 =\lim_{N\to \infty}\CW_{\rm U}(J.J^\dagger,N)/N^2$ 
was shown to satisfy a partial differential equation with respect to the
eigenvalues of $J.J^\dagger$ \cite{BG}. In the
``strong coupling phase'', an explicit expression was given \cite{OBZ} for the 
expansion of $W_\rU$ in a series expansion in traces of powers of
$J.J^\dagger$
\be
W_{\rm U}(J.J^\dagger)= \sum_{n=1}^\infty \sum_{\Ga\,\vdash n}
W_\Ga\, \frac{\ttr_\Ga J.J^\dagger}{\prod_p\left(\Ga_p! \, p^{\Ga_p}\right)} 
\ee
\be
\qquad W_\Ga = (-1)^n {(2n+\sum \Ga_p -3)!\over (2n)!}
\prod_{p=1}^n \left({-(2p)!\over p! (p-1)!}\right)^{\Ga_p}\ , 
\ee
where $\Ga\vdash n$ denotes a partition of $n=\Ga_1.1+\Ga_2.2+\cdots+\Ga_n.n$
and 
\be 
\ttr_\Ga(X):= \prod_{p=1}^n \(\frac{1}{N} \tr X^p\)^{\Ga_p}\ .
\ee

Now we claim that $W_{\rm O}$ defined in (\ref{WOinf}) is
\be\label{claim}
  W_{\rm O}(J.J^t) ={1\over 2}W_{\rm U}(J.J^t)\ .
\ee

{\it Proof}. We repeat the steps of \cite{BG}, paying due attention 
to the differences between independant matrix elements in a complex
Hermitian and in a real symmetric matrix. The trivial identity
$\sum_j \frac{\partial^2 \CZO}{\partial J_{ij}\partial J_{kj}}=N^2 \delta_{ik} \CZO$ is reexpressed in terms of the eigenvalues $\Gl_i$
of the real symmetric matrix  $J.J^t$:
\be 
4 \Gl_i \frac{\partial^2\CZO}{\partial \Gl_i^2} 
+2\sum_{j\ne i} \frac{\Gl_j}{\Gl_j-\Gl_i}\left(\frac{\partial \CZO}{\partial\Gl_j}
-\frac{\partial \CZO}{\partial\Gl_i}\right) +2N\frac{\partial \CZO}{\partial\Gl_i}
=N^2 \CZO\ .
\ee
Writing as above $\CZO=\rme^{N^2 W_{\rm O}}$ and dropping subdominant terms
in the large $N$ limit, with $W_{\rm O}$ and $W_i:=N {\partial W_{\rm O}}/{\partial\Gl_i}$ of
order 1, we get
\be
4\Gl_i W_i^2 + 2W_i +\frac{1}{N}\sum_{j\ne i} \frac{2\Gl_i}{\Gl_j-\Gl_i}(W_j-W_i)=0
\ee
which is precisely the equation satisfied by $\case{1}{2} W_{\rm U}$ 
in \cite{BG}. This, supplemented by appropriate boundary conditions, suffices 
to complete the proof of (\ref{claim}).

As noticed elsewhere \cite{IZ,ZJZ} it is appropriate to expand $W$ on ``free''
(or ``non-crossing'') cumulants \cite{BIPZ,IZ,Spe}
\be
\psi_q(X):=-\sum_{\Ga_1,\ldots,\Ga_q\ge 0\atop\sum_i i\Ga_i=q} 
\frac{(q+\sum_i \Ga_i-2)!}{(q-1)!} 
\prod_i \frac{(-\frac{1}{N}\tr X^i)^{\alpha_i}}{\Ga_i!}\ .
\ee
Then
\be
\!\!\!\!\!\!\!\!\!\!\!\!\!\!\!\!\!\!\!\!\!
W_{\rm O}(J.J^t) ={1\over 2} \psi_1(J.J^t)-{1\over 4}\psi_2(J.J^t) +{1\over 3}\psi_3(J.J^t)
-{1\over 8} (5\psi_4(J.J^t)+\psi_2^2(J.J^t))+\cdots
\ee
where the only occurrence of $\psi_1$ is in the first term. 

\section{The generalized HCIZ integral}

\subsection{Notations. Review of known results}

Let us first recall well known results. If the matrices $A$ and $B$ in 
(\ref{Theint}) are in the Lie algebra of the group $G$, namely 
are  real skew-symmetric,  respectively anti-hermitian, for $G$ = O($N$), 
resp. U$(N)$, the exact  expression of  $Z^{(G)}$ is known from the
work of Harish-Chandra \cite{HC}. To make the formulae quite explicit, we 
take  $A$ and $B$ in the Cartan torus, i.e. of a diagonal or block-diagonal
 form~\cite{PFEDFZ} 
\be 
A=\diag\left( a_i\right)_{i=1,\cdots,N} \qquad {\rm for\ U}(N)
\ee  
and
\be 
\label{CartanO}
A= \cases{\diag\left(\pmatrix{0&a_i\cr -a_i&0}_{i=1,\cdots,m} \right)
& for O($2m$)\cr  
 \diag\left(\pmatrix{0&a_i\cr -a_i&0}_{i=1,\cdots,m},0 \right)  
& for O($2m+1$)\cr }
\ee
and likewise for $B$. We assume that all the $a$'s are distinct, and likewise
for the $b$'s. 
Then the integral (\ref{Theint}) reads 
\be
\label{HCresult}
Z^{(G)}={\rm const.}\, \frac{\det(\CM_G)}{\Delta_G(a)\Delta_G(b)}
\ee
where
\be
\label{HCdetail}
\hskip -30mm
\CM_G=\cases{\left(\rme^{\kappa N a_i b_j}\right)_{1\le i,j\le N} \cr \cr
\left(\rme^{2 \kappa N a_i b_j}+\rme^{-2\kappa N a_i b_j}\right)_{1\le i,j\le m} \cr\cr
\left(\rme^{2\kappa N a_i b_j}-\rme^{-2 \kappa N a_i b_j}\right)_{1\le i,j\le m}\cr
}
\ ,\quad 
\Delta_G(a)=\cases{\prod_{i<j} (a_i-a_j) & if $G=$ U($N$)\cr
\prod_{1\le i<j\le m}\!\!\!\! (a^2_i-a^2_j) & if $G= \rO(2m)$\cr
\prod_{1\le i<j\le m}\!\!\!\! (a^2_i-a^2_j)\prod_{i=1}^m a_i & if $G=\rO(2m+1)$\cr
}
\ee
For convenience, $\Delta_\rU(a)$ 
will be abbreviated into $\Delta(A)=\prod_{i<j} (a_i-a_j)$,  
the usual Vandermonde determinant.


\subsection{The large N limit}
We now consider 

\be 
\label{FO}
F^{(\rO)}(A,B):= \lim N^{-2} \log \int DO \,\rme^{N \tr A O B O^t}
\ee

\be
\label{FU}
F^{(\rU)}(A,B):= \lim N^{-2} \log \int DU\, \rme^{ {\bf 2} N \tr A U B \Ud}\ .
\ee
Note the factor 2 introduced for convenience in the latter exponential. 
We claim that, for $A$ and $B$ both symmetric or skew-symmetric, 
\be
\label{claim2}
 F^{(\rO)}(A,B)=\oh F^{(\rU)}(A, B)\ .
\ee

\subsubsection{The skew-symmetric case.}

For $A$ and $B$ both skew-symmetric, the limit of (\ref{HCresult})
is easy to evaluate.
Assuming without loss of generality that all $a$'s and $b$'s
are positive, we find that 
$(\CM_\rO)_{ij}\sim \rme^{2 N a_i b_j}$, thus 
$Z^{(\rO)}\sim \det \left( \rme^{2 N a_i b_j} \right)/\Delta_\rO(a)
\Delta_\rO(b)$.
On the other hand, real matrices $A$ and $B$ of the form (\ref{CartanO})
may be also regarded as antiHermitian, with eigenvalues 
$A_j=\pm \rmi a_j$, ${j=1,\cdots,m}$ (supplemented by $0$ if $N=2m+1$).
An easy computation gives $\Delta(A)=(\Delta_\rO(a))^2$ up a to sign. 
The matrix $\CM_\rU=(\rme^{2 N  A_i B_j})_{1\le i,j\le N}\approx
(\rme^{2N a_i b_j})_{1\le i,j\le m}\otimes \pmatrix{0&1\cr 1&0}$, 
as $N\to \infty$. Hence $\det \CM_\rU \sim (\det \CM_\rO)^2$.
Thus the $\rU(N)$ integration for the pair $(A, B)$ 
yields, according to (\ref{HCdetail}), and up to an overall 
factor 
\be
\hskip -20mm
\ZU(A,B)= {\det \left( \rme^{2 N A_i B_j} 
\right)\over\Delta(A)\Delta(B)}\sim   \({(\det (\rme^{2N a_i b_j})_{1\le i,j\le m}\over \Delta_\rO(a)\Delta_\rO(b)}\)^2
={\rm const.}\, (Z^{(\rO)(A,B)})^2 
\ee
in accordance with (\ref{claim2}).

\subsubsection{The symmetric case.}

We now turn to the more challenging case where both $A$ and $B$ 
are real symmetric. We may suppose that $A$ and $B$ are in a diagonal form
$A=\diag(a_i)$, $B=\diag(b_i)$, and assume that
 all $a$'s and all $b$'s are distinct. 
In that case, we shall resort to (an infinite set of)
differential equations, in a way similar to the discussion of sect. 2.

In a recent work,  Berg\`ere and Eynard \cite{BE} have introduced the 
following integrals over the compact group $G$
\be 
\label{bergeyn}
M_{ij}=
\int_{G} d\Omega\, \Omega_{ij}\Omega^{-1}_{ji}
 \rme^{\kappa N \tr A\GO B\GO^{-1}}\ ,
\ee
which may be regarded as particular two-point correlation 
functions associated with 
 the partition function $Z^{(G)}$ of (\ref{Theint}). 
The latter is recovered from $M_{ij}$ by a summation 
over $i$ or $j$
\be 
Z^{(G)}=\sum_i M_{ij},\  \forall j, \qquad 
 Z^{(G)}=\sum_j M_{ij},\ \forall i\ .
\ee
As shown in \cite{BE}, $M_{ij}$ satisfy the following set 
of differential equations
\be \label{diffeqM}
\sum_j K_{ij} M_{jk} = \kappa N M_{ik} b_k
\ee
with  no summation over $k$ in the r.h.s. Here $K$ refers to 
the matrix differential operator
\be
\label{Kdiff}
K_{ii}=\frac{\partial}{\partial a_i}+ \frac{\beta}{2}\sum_{j\ne i}
\frac{1}{a_i-a_j}\qquad {\rm and\ for \ } i\ne j,\quad K_{ij}=
- \frac{\beta}{2}\frac{1}{a_i-a_j}\ .
\ee
(We  make use of Dyson's label $\beta=1, 2$ 
for  $G=$ O($N$), U($N$)
, respectively.)
  
Now, by repeated application of the operator $K$ on $M$,  
we find for any positive integer $p$ that
$\sum_j (K^p)_{ij} M_{jk} = N^p M_{ik} b_k^p$, hence, after 
summation over $i$ and $k$
\be 
\sum_{i,j} (K^p)_{ij}\,  Z^{(G)}
 =Z^{(G)} \sum_k (\kappa N b_k)^p\ .
\ee
The differential operator $D_p:= \sum_{1 \le i,j\le N} (K^p)_{ij}$
has thus the property that
\be 
\label{diffp}
 D_p  Z^{(G)}
=(\kappa N)^p \tr B^p\,  Z^{(G)}\ .
\ee
Thus far, the discussion holds for any finite value of $N$.
Now take the large $N$ limit with the Ansatz $  Z^{(G)}
=\rme^{N^2 F^{(G)}}$. 
Equation (\ref{diffp}) reduces in that limit to 
\be
\label{diffF}
\sum_i \({N\over\kappa} {\partial F^{(G)}\over \partial a_i}+\frac{\beta}{2\kappa N}
\sum_{j\ne i}\frac{1}{a_i-a_j}\)^p= \tr B^p
\ee
with $  N {\partial F\over \partial a_i}$ of order 1, as in section 2, 
and all other terms resulting from further application 
of $\partial/\partial_{a_i}$ over $ F^{(G)}$ suppressed by inverse powers 
of $N$. 

These equations had been obtained in \cite{IZ} in the case of $\rU(N)$
($\beta=2$) from the explicit form of $Z^{(\rU)}$ and shown 
to determine recursively the expansion of $\FU$ in 
traces of powers of $A$ and $B$. Comparing the orthogonal
($\kappa=\beta=1$) and unitary ($\kappa=\beta=2$) cases, it is clear on (\ref{diffF})
that  $2 F^{(\rO)}(A,B)$  satisfies the same set of equations
as $F^{(\rU)}(A,B)$, thus vindicating (\ref{claim2}).

\ommit{\item For $A$ and $B$ both symmetric, 
the validity of (\ref{claim2}) has been tested up to order 6 
in an expansion in powers of $A$ or $B$.}


\subsection{The generic case}
Although it does not look very natural in view of their
symmetries, one may extend the integrals $\ZU$ and $\ZO$ 
to the case of generic complex (non Hermitian), resp. real
(neither symmetric nor skew-symmetric),  matrices $A$ and $B$.
If we insist on having real quantities in the exponential, the 
unitary integral that we consider reads
\be 
\ZU=\int dU \rme^{\tr (A U B U^\dagger + A^\dagger U B^\dagger U^\dagger)}
\ee
(and one recovers the factor 2 introduced in (\ref{FU}) 
for $A$ and $B$ Hermitian).
In parallel the more general orthogonal integral reads
\be
\ZO=\int dO \rme^{\tr A O B O^t}=\int dO  \rme^{\tr A^t O B^t O^t}
\ee
The functions $\ZU$ and $\FU$ 
have now expansions in traces of products of $A$ and $A^\dagger$
(or $A^t$)
and of $B$ and $B^\dagger$ (or $B^t$), with an equal number of daggers 
(resp. transpositions) appearing
on $A$ and $B$. We can no longer rely on the diagonal form 
of $A$ and $B$ (a generic real, resp. complex matrix is {\it not}
diagonalized by an orthogonal, resp. unitary, matrix) and there 
are no longer  differential equations in these eigenvalues
satisfied by $Z$ or $F$. Still, there is some evidence that 
universality  holds again. By expanding the exponentials
and by making use of the explicit integrals (\ref{basic}), (\ref{basicU}),
(see also Appendix A), 
we have checked that, up to 4th order, for $A$ and $B$ real
\be
\label{claim3}
\FO(A,A^t; B,B^t)=\oh \FU(A,A^\dagger=A^t; B, B^\dagger=B^t)
\ee

If we write the expansion of $F$ in powers of $A$ (and $B$) as
$F=\sum_{n=1} F_n$, we find
 \be \FU_1= 
\psi_1(A) \psi_1(B)+\psi_1(\Ad) \psi_1(\Bd)
 \ee

\be \FU_2=\oh \( \psi_2(A)\psi_2(B) +2 \psi_{2t}(A,\Ad)\psi_{2t}(B,\Bd)+
\psi_2(\Ad)\psi_2(\Bd) \) \ee

\be \! \! \! \! \! \FU_3=\frac{1}{3}\( \psi_3(A)\psi_3(B)+3  \psi_{3t}(A,\Ad)\psi_{3t}(B,\Bd)
+(A\to \Ad,B\to \Bd) \)\ee
\begin{eqnarray}
\! \! \! \! \! \! \! \! \! \! \! \! \! \! \! \! \! \! \! \! \! \! \! \! \! \! \! \! \! \! \! \! \! \! \! \! \! \! \! \! \! \! 
\FU_4 = \frac{1}{4} \(
\psi_4(A\right.)&\psi_4(B) +4\psi_{4t}(A,\Ad)\psi_{4t}(B,\Bd)+2 \psi_{4tt}(A,\Ad)\psi_{4tt}(B,\Bd)
\cr
& \! \! \! \! \! \! \! \! \! \! \! \! \! \! \!  \! \! \! \! \! \! \! \! \! \! \! \! \! \! \! 
+ \psi_{4t|t}(A,\Ad)\psi_{4t|t}(B,\Bd)- \psi^2_2(A)\psi^2_2(B)- \psi^2_{2t}(A,\Ad)\psi^2_{2t}(B,\Bd)
-\psi^2_{2t}(A,\Ad)\psi_{2}(B)\psi_{2}(\Bd)
\cr
& 
 \! \! \! \! \! \! \! \! \! \! \! \! \! \! \!  \! \! \! \! \! \! \! \! \! \! \! \! \! \! \! \left.
-\psi_{2}(A)\psi_{2}(\Ad)\psi^2_{2t}(B,\Bd)
-4 \psi_{2}(A) \psi^2_{2t}(A,\Ad)\psi_{2}(B)\psi^2_{2t}(B,\Bd) +(A\to \Ad,B\to \Bd)
\) 
\end{eqnarray}
where the $\psi_n$ are the free cumulants defined above 
and the 
$\psi_{nt}$ are ``polarized'' versions of the latter, involving 
$A$ and $\Ad$ (or $A^t$), see Appendix B for explicit expressions.

\section{Concluding remarks}

\begin{itemize}

\item
 It should be stressed that the equality (\ref{claim2})
can be true only asymptotically 
as $N\to \infty$. Indeed the exact result \cite{HC} for $A$ and $B$ skew-symmetric
as well as what is known for $A$ and $B$ both symmetric \cite{BH} clearly 
indicate that it does not hold for finite $N$.

\item  The differential operator $D_p$ considered above is 
interesting in its own right. 
Consider the differential operator $\hat D_p(\partial /\partial A)$
such that 
\be 
\label{ppty}
\hat D_p(\partial /\partial A)\, \rme^{N\tr A B}=N^p \tr B^p \,\rme^{\tr A B}
\ee
For Hermitian matrices, for which all matrix elements $A_{ij}$ may
be regarded as independent, one may write
\be
\label{Dtr}
 D_p(\partial /\partial A) =
\tr\(\frac{\partial}{\partial A}\)^p := \sum_{i_1,\cdots i_p}
\frac{\partial}{\partial A_{i_1 i_2}}\frac{\partial}{\partial A_{i_2 i_3}}
\cdots \frac{\partial}{\partial A_{i_p i_1}}
\ee
while in the case of symmetric matrices, the general expression 
involves some combinatorial factors. 
The above property (\ref{ppty}) suffices to define $\hat D_p$ on any (differentiable)
function of $A$, by Fourier transform. 

Now let $\hat D_p$ act on functions  $f(A)$ invariant upon 
$A\to \Omega A \Omega^{-1}$. Then $\hat D_p$ reduces to a 
differential operator $D_p$ on the eigenvalues $a_i$ of $A$.
As we have seen above, $D_p=\sum_{ij}(K^p)_{ij}$ but it 
would seem desirable to have a more direct construction of
that basic operator.
In the case of $G=\rU(N)$, one has  the elegant form  \cite{IZ}
\be
\label{expl}
 D_p=\frac{1}{\Delta(A)} \sum_i \({\partial\over\partial a_i}\)^p
\Delta(A)\ .
 \ee
This result, however, makes use of the explicit form  (\ref{HCresult}) of
 $Z^{(\rU)}$, and there is no counterpart for $G=\rO(N)$. 
 Thus the question is :
Can one derive the expression  (\ref{expl}) of $D_p$ from 
that (\ref{Dtr}) of $\hat D_p$? Curiously, what looks like an innocent 
exercise of calculus turns out to be non trivial, even for 
$G=\rU(N)$.

\item In view of the similarity between (\ref{claim}) and (\ref{claim2}), 
on the one hand, and of our (partial) results 
and conjecture on the `generic' case, on the other,  it would be nice to have 
a general, intuitive argument why these universality properties
hold. Heuristically, the overall factor $\oh$ 
in (\ref{claim})  and (\ref{claim2}) just reflects the 
ratio of numbers of degrees of freedom  in the two cases:
there are $N(N-1)/2\sim N^2/2$ real parameters in an orthogonal
matrix, and $N^2$ in a unitary one. But why is the function of $A$ and $B$
universal?

\item {\it Diagrammatics?}  
A diagrammatic expansion exists for $\FU$ \cite{ZJZ}, using the 
functional $\CW$ of section 2, and this matches 
a combinatorial expansion \cite{Collinsth}. 
Repeating the argument  in the real orthogonal case
leads to a much less transparent result, however, and does not seem to yield a 
simple derivation of (\ref{claim2}) based on (\ref{claim}).

\item {\it Historical remarks}. 
As far we know, the property (\ref{claim}) had never 
been observed before. On the other hand, property (\ref{claim2})
has a richer history. It seems to have been first observed in the
case where $A$ or $B$ is of finite {\it rank} in \cite{MPR}, 
and then repeatedly used in the physics literature \cite{XX,Bouchaud}. 
This was later proved in a rigorous way in \cite{Collinsth}.
In \cite{CS2}, this is extended to the case where  the rank is o($N$).
Indeed for finite rank of $A$, say,
 only terms with a single trace of some power of $A$ dominate, and
the expression of $F(A,B)$ is known to be given 
by $\sum_{n\ge 1} \case{1}{n} \case{1}{N}\tr A^n\psi_n(B)$ for 
the unitary group \cite{IZ}.

Following a totally different approach, Guionnet and Zeitouni \cite{GZ}
have proved rigorously the existence of the free energies $\FU$ and
$\FO$ (for $A$ and $B$ symmetric)  
 in the large limit, and have established that they solve the
flow equation proposed by Matytsin \cite{Ma}. A byproduct of their 
discussion is the explicit $\beta$ dependence of the free 
energy and the resulting universality property (\ref{claim2}).

\ommit{For $A$ and $B$ diagonal (or just symmetric)  
and $A$ of finite {\it rank} \cite{Parisi}, or more generally of
rank o($N$) \cite{CS2}, only terms with a single trace of $A$ dominate, and
the expression of $F(A,B)$ is known to be given 
by $\sum_{n\ge 1} \case{1}{n} \case{1}{N}\tr A^n\psi_n(B)$ for 
the unitary group 
\cite{IZ} and by  $\sum_{n\ge 1} \case{2^{n-1}}{n} \case{1}{N}\tr A^n\psi_n(B)$
for the orthogonal group.  The latter was anticipated in \cite{Parisi},
 repeatedly used in the physics literature \cite{XX,Bouchaud} and rigorously 
proved in \cite{CS2}.}

\end{itemize}

\ack{The author was supported by
``ENIGMA'' MRT-CT-2004-5652,
ESF program ``MISGAM'' and ANR program ``GIMP'' ANR-05-BLAN-0029-01.
He wants to  thank A. Guionnet and P. Zinn-Justin for discussions, 
and Y. Kabashima for revigorating his interest in these integrals.
Special thanks go to M. Berg\`ere and B. Eynard for communicating 
their results prior to publication. }

\section*{Appendix A. More details on the `basic integrals'}

In this Appendix, we recall well known results \cite{We} 
on the integral (\ref{basic0}). Equivalently we may consider 
\be{\cal I}(\bu,\bv)=\int DO \prod_{a=1}^n (\bu_a. O\bv_a)\ ,\ee
where $\bu_a$ and $\bv_a$, $a=1,\cdots, n$,  are  vectors of $\Bbb{R}^N$.  
The integral ${\cal I}(\bu,\bv)$ is linear in each  $\bu_a$ and each 
$\bv_a$, 
and is invariant under a global rotation of all $\bu$'s or of all $\bv$'s:
$\bu_a\to O_1 \bu_a$, $\bv_a \to O_2 \bv_a$, since this may be absorbed by
the change of integration variable $O_1^t O O_2 \to O$ in accordance with
 the invariance of the Haar measure $DO$.  If $N> n$ the 
completely antisymmetric tensor $\epsilon_{}$ cannot 
be used to build invariants. Hence 
${\cal I}(\bu,\bv)$ is only function of the invariants 
$\bu_a.\bu_b$, $\bv_a.\bv_b$ and by linearity must be of the form
\be
{\cal I}(\bu,\bv)=\sum_{p_1,p_2}C(p_1,p_2) \prod (\bu_a.\bu_{p_1(a)})
\prod(\bv_b.\bv_{p_2(b)})\ ,
\ee
a sum over all possible {\it pairings} of the indices $a=1,\cdots,n$, 
$b=1,\cdots,n$; this shows that ${\cal I}$ vanishes for $n$ odd. In the following 
we change $n\to 2n$ and denote $P_{2n}$ the set of all pairings of 
$\{1,2,\cdots, 2n\}$, with $|P_{2n}|=(2n-1)!!$ . 

Then the general expression of $\int DO\, O_{i_1j_1}\cdots O_{i_{2n}j_{2n}}$
is indeed of the form (\ref{basic}). 
The coefficients  $C(p_1,p_2)$  
may be determined recursively, 
but let us first point some general features. 

\noindent (i) Regard now  $p_1$ and $p_2$  as  permutations of $S_{2n}$, both 
in the class $[2^n]$ of permutations made of $n$ 2-cycles (transpositions). 
Represent a  typical term in the r.h.s. of (\ref{basic})
by a set of disjoint
chain loops $i_a - j_a - j_{p_2(a)}-  i_{p_2(a)}- i_{p_1.p_2(a)}- \cdots$,
(these are the loop diagrams of \cite{We}). 
The  coefficients $C(p_1,p_2)$ are thus only functions of the 
product  $p_1.p_2$, and in fact functions only of the {\it class} in 
$S_{2n}$ of that product. Indeed if 
all $i$ and $j$ indices are relabelled
through the same permutation $\pi\in S_{2n}$, 
$i_a\to i'_a=i_{\pi(a)}$, 
$j_a\to  j'_a=j_{\pi(a)}$, 
$a=1,\cdots, 2n$, 
the integrand is preserved and 
 $p_s\to p_s'=\pi^{-1}. p_s.\pi$, 
for $s=1,2$, hence $p_1.p_2 \to \pi^{-1}. p_1.p_2.\pi$
and $C(p_1.p_2)$ must depend only on the class $[p_1.p_2]$.

\noindent
(ii) For $p_1$ and $p_2 \in [2^n]$, their product 
$p_1. p_2$ is the product of two permutations of $S_n$ acting on
two disjoints subsets of 
$n$  elements of $\{1,2,\cdots,2n\}$, both in
the same class of $S_n$, $p_1. p_2=\sigma.\sigma'$ with
$[\sigma]=[\sigma']$  \cite{CS2}. The class $[p_1. p_2]$ of $p_1. p_2$ is 
completely specified by $[\sigma]$, hence we may write  the coefficients 
as $C(p_1,p_2)=C([\sigma])$. 

\noindent{\it Proof.} To any cycle $\Ga$ of $p_1.p_2$, 
$\{ a, p_1.p_2(a), (p_1.p_2)^2(a),\cdots, (p_1.p_2)^r(a)\}$, 
one may associate another one
$\{ p_1(a), p_1.p_2.p_1(a), (p_1.p_2)^2p_1(a),\cdots, (p_1.p_2)^rp_1(a)\}$,
which is obviously of the same length and which acts on distinct elements. 
Thus $p_1.p_2= \sigma.\sigma'$, where   $ \sigma$ and $\sigma'$ acting on 
distinct elements of  $\{1,2,\cdots,2n\}$ may be regarded as in the same class
of  $S_n$.
Moreover the class $[p_1.p_2]$, i.e. the cycle structure of $p_1.p_2$ is 
obviously given by that of $[\sigma]=[\sigma']$. q.e.d.
\\
The coefficients $C$ are then determined recursively. 
Noting that by contracting the last two $j$ indices one 
constructs $O_{i_{n-1}j_{n-1}}O_{i_{n}j_{n}}\delta_{j_{n-1}j_n}=(O.O^t)_{i_{n-1}i_n}=
\delta_{i_{n-1}i_n}$, one gets a (strongly overdetermined) system of 
equations relating the $C$'s of order $n$ to those of order $n-1$ \cite{We}.
Explicit although fairly complicated solutions have been given \cite{CS1}.

\def\di#1#2{\delta_{i_{#1}i_{#2}}}\def\dj#1#2{\delta_{j_{#1}j_{#2}}}

\medskip
The first coefficients read explicitly 
$$
\eqalign{n&=1\quad \scriptstyle{C[1]={1\over N}}\cr
n&=2\quad \scriptstyle{C[2]=\frac{-1}{N (N-1)(N+2)}}\ ,\quad\scriptstyle{C[1,1]=\frac{N+1}{N (N-1)(N+2)}}\cr
n&=3\quad \scriptstyle{C[3]=\frac{2}{(N-2) (N-1) N (N+2) (N+4)}\ ,\quad
C[1,2]=\frac{-1}{(N-2) (N-1) N (N+4)}\ , \quad
C[1^3]=\frac{N^2+3 N-2}{(N-2) (N-1) N (N+2) (N+4)}}
\cr
n&=4 \quad\scriptstyle{C[4]=\frac{-(5 N+6)}{(N-3) (N-2) (N-1) N (N+1) (N+2) (N+4) (N+6)}\ ,\quad
C[1,3]=\frac{2}{(N-3) (N-2) (N-1) (N+1) (N+2) (N+6)}}\cr
&\qquad\quad  
 \scriptstyle{C[2^2]=\frac{N^2+5 N+18}{(N-3) (N-2) (N-1) N (N+1) (N+2) (N+4) (N+6)}\,,\
C[1^2,2]=\frac{-(N^3+6 N^2+3 N-6)}{(N-3) (N-2) (N-1) N (N+1) (N+2) (N+4) (N+6)}\,,}\cr
&\qquad\quad  \scriptstyle {C[1^4]=\frac{(N+3) \left(N^2+6 N+1\right)}{(N-3) (N-1) N (N+1) (N+2) (N+4) (N+6)} }\cr
%
}$$

The analogous basic integrals on $\rU(N)$ are more widely known, 
see (\ref{basicU}).
One may actually give an explicit form to the $C([\sigma.\tau])$, 
namely
$$\int DU\, U_{i_1j_1}\cdots U_{i_nj_n} \Ud_{k_1\ell_1} \cdots\Ud_{k_n\ell_n}=
\sum_{\tau,\sigma\in S_n}
\sum_{Y {\rm Young\ diagr.}\atop |Y|=n}
{(\chi^{(\lambda)}(1))^2 \chi^{(\lambda)}([\sigma])\over n!^2 s_\lambda(I)}
\prod_{a=1}^n \delta_{i_a\ell_{\tau(a)}} \delta_{j_a k_{\tau\sigma(a)}}\ .$$
where $\chi^{(\lambda)}([\sigma])$ is the character of the symmetric
group $S_n$ associated with the Young diagram $Y$, hence a function of the class
 $[\sigma]$ of $\sigma$~;  $\chi^{(\lambda)}(1)$ is thus the
dimension of that representation~; $s_\lambda(X)$ is the character
of the linear group GL($N$) associated with the Young diagram $Y$, 
i.e. a Schur function when expressed in terms of the eigenvalues of
$X$~; $s_\lambda(I)$  is thus the dimension of that representation.

The first coefficients read explicitly
$$\eqalign{n&=1\quad \scriptstyle{C[1]=\frac{1}{N}}\cr
n&=2\quad
\scriptstyle{C[{2}]=-\frac{1}{(N - 1) N (N + 1)}\ ,\quad
C[{1, 1}]=\frac{1}{(N - 1) (N + 1)}}\cr
n&=3\quad
\scriptstyle{C[{3}]=\frac{2}{(N -  2) (N - 1) N (N + 1) (N + 2)}
\ ,\quad
C[{2,    1}]=-\frac{1}{(N - 2) (N - 1) (N + 
            1) (N + 2)}
\ ,\quad
C[{1^3}] =\frac{N^2 -         2}{(N - 2) (N - 1) N (N + 1) (N + 2)}
}\cr
n&=4\quad
\scriptstyle{
C[{4}]=-\frac{5}{(N - 3) (N - 2) (N - 1) N (N + 1) (N + 2) (N + 3)}
\ ,\quad
C[{3,  1}]=\frac{2 N^2 - 
        3}{(N - 3) (N - 2) (N - 1) N^2 (N + 1) (N + 2) (N + 3)}
}\cr
&\scriptstyle{C[{2^2}] 
=\frac{N^2 + 
        6}{(N - 3) (N - 2) (N - 1) N^2 (N + 1) (N + 2) (N + 3)}
}\,,\ \
C[{2,1^2}] 
=-\frac{1}{(N - 3) (N - 1) N (N +      1) (N + 3)}
\,,\ \
C[{1^4}] =\frac{N^4 - 8 N^2 + 
        6}{(N - 3) (N - 2) (N - 1) N^2 (N + 1) (N + 2) (N + 3)}
}$$

\section*{Appendix B. Free (non crossing) cumulants}
 Beware ! In this Appendix, we make use of a different notation for 
normalized traces
$\ttr X:=\1oN \tr X$, $\tr X$ the usual trace, thus $\ttr I =1$.

For convenience, we list here the first free cumulants of $A$
in terms of the $\phi_p(A)= \ttr A^p= \1oN \tr A^p$ 
together with the mixed ones, involving traces of products $A$ and $\Ad$.

\ommit{$$ \psi_m=-\sum_{\alpha\vdash m}{(m+\sum \alpha_p -2)!\over (m-1)!}
\prod_p {(-\phi_p)^{\alpha_p}\over \alpha_p!}$$
or more explicitly 
$$\eqalign{
\phi_1 &=\psi_1\cr
\phi_2 &=\psi_2+\psi_1^2\cr
\phi_3 &=\psi_3+3 \psi_1\psi_2+\psi_1^3\cr
\phi_4 &=\psi_4+4\psi_1\psi_3 +2\psi_2^2+6 \psi_1^2\psi_2+\psi_1^4\cr}$$}

$$\eqalign{
\psi_1(A) &= \ttr A \cr 
\psi_2(A) &=\ttr A^2 -(\ttr A)^2 \cr 
\psi_3(A) &=\ttr A^3 -3 \ttr A \ttr A^2 +2 (\ttr A)^3 \cr 
\psi_4(A) &= \ttr A^4 -4\ttr A \ttr A^3 -2(\ttr A^2)^2 +10 (\ttr A)^2 \ttr A^2 -5 (\ttr A)^4
\cr}$$

$$\eqalign{
\psi_{2t}(A,\Ad)&=\ttr(A \Ad) - \ttr{A}\ttr \Ad \cr
\psi_{3t}(A,\Ad)&=   \ttr( A^2 \Ad) - \ttr {\Ad}\,\ttr(A^2) - 2\ttr{A}\,\ttr(A\Ad) +  2(\ttr{A})^2\ttr\Ad\cr
\psi_{4t}(A,\Ad)&=\ttr (A^3\Ad) - \ttr \Ad \ttr A^3-3 \ttr A \ttr (A^2 \Ad)) 
-2 \ttr A^2 \ttr(A \Ad)+ 5 \ttr A\ttr \Ad \ttr(A^2)\cr 
  &\qquad +5(\ttr A)^2\ttr(A \Ad))-5 \ttr^3 A\ttr \Ad\cr
\psi_{4tt}(A,\Ad)&=
\ttr (A^2 \Ad^2) - 2\ttr \Ad \ttr (A^2 \Ad) - 2\ttr A \ttr (A \Ad^2) 
- \ttr(A^2)\ttr(\Ad^2) -(\ttr(A \Ad))^2  \cr
& \qquad + 2 (\ttr A)^2 \ttr(\Ad^2) + 2 (\ttr \Ad)^2 \ttr(A^2)
 +6\ttr A \ttr \Ad\ttr(A \Ad)-5 \ttr^2 A \ttr^2 \Ad\cr
\psi_{4t-t}(A,\Ad)&=
\ttr ((A \Ad)^2) - 2\ttr \Ad \ttr (A^2 \Ad) - 2\ttr A \ttr (A \Ad^2) 
 -2(\ttr(A \Ad))^2 \cr
& \qquad + (\ttr \Ad)^2 \ttr A^2 + (\ttr A)^2 \ttr \Ad^2 
+8\ttr A \ttr \Ad \ttr(A \Ad))-5 (\ttr A)^2 (\ttr\Ad)^2\ .\cr
}$$

\end{document}